\documentclass[aps,prl,twocolumn,superscriptaddress]{revtex4-2}
\usepackage{mathtools}
\usepackage{xcolor}
\usepackage{graphicx}
\usepackage{amsmath}
\usepackage{amssymb}

\newcommand{\e}{\varepsilon}              %

\begin{document}

\title{Internal reliability and anti-reliability in dynamical networks}

 \author{Tommaso Matteuzzi}
 \email{tommaso.matteuzzi@unifi.it}
 \affiliation{Department of Physics and Astronomy  and CSDC, University of Firenze, via G.Sansone 1, I-50019, Sesto Fiorentino, Italy}
 \author{Franco Bagnoli}
 \email{franco.bagnoli@unifi.it}
 \affiliation{Department of Physics and Astronomy and CSDC, University of Firenze, via G.Sansone 1, I-50019, Sesto Fiorentino, Italy}
 \email{franco.bagnoli@unifi.it}
 \affiliation{Istituto Nazionale di Fisica Nucleare, sezione di Firenze, via G.Sansone 1, I-50019, Sesto Fiorentino, Italy}
 \author{Michele Baia}
 \email{michele.baia@unifi.it}
 \affiliation{Department of Physics and Astronomy  and CSDC, University of Firenze, via G.Sansone 1, I-50019, Sesto Fiorentino, Italy}
 \affiliation{Istituto Nazionale di Fisica Nucleare, sezione di Firenze, via G.Sansone 1, I-50019, Sesto Fiorentino, Italy} 
 \author{Stefano Iubini}
 \email{stefano.iubini@cnr.it}
 \affiliation{Istituto dei Sistemi Complessi, Consiglio Nazionale delle Ricerche, via Madonna del Piano 10, I-50019, Sesto Fiorentino, Italy} 
  \affiliation{Istituto Nazionale di Fisica Nucleare, sezione di Firenze, via G.Sansone 1, I-50019, Sesto Fiorentino, Italy}
\author{Arkady Pikovsky}
 \email{pikovsky@uni-potsdam.de}
 \affiliation{Department of Physics and Astronomy, University of Potsdam, Karl-Liebknecht-Str. 24/25, 14476, Potsdam-Golm, Germany}

\date{\today}

\begin{abstract}
We consider finite dynamical networks and define internal reliability according to the synchronization properties of a replicated unit or a set of units. If the states of the replicated units coincide with their prototypes, they are reliable; otherwise, if their states differ, they are anti-reliable. Quantification of reliability with the transversal Lyapunov exponent allows for a straightforward analysis of different models. For a Kuramoto model of globally coupled phase oscillators with a distribution of natural frequencies, we show that prior to the onset of synchronization, peripheral in frequency units are anti-reliable, while central are reliable. For this model, reliability can be expressed via phase correlations in a sort of a fluctuation-dissipation relation. Sufficiently large sub-networks in the Kuramoto model are always anti-reliable; the same holds for a recurrent neural network, where individual units are always reliable.  
\end{abstract}

\maketitle

\textit{Introduction.}  The stability properties of dynamical states lie at the core of characterizing nonlinear dynamical systems. Sensitive dependence of the dynamics on initial conditions, quantified by positive Lyapunov exponents (LEs), serves as a definition of chaos~\cite{Ott-book-02}. In chaotic systems, in the course of the dynamics, memory on the initial state disappears. If one prepares two identical systems (two replicas) but the initial conditions are slightly different, then the states of the replicas will not be correlated after some transient. The notions of reliability and anti-reliability~\cite{Mainen-Sejnowski-95,Goldobin-Pikovsky-06b,ermentrout2008reliability,lin2009reliability} are used to describe the response of replicas to nontrivial external forces, in particular to external noise. Here, if the largest LE (LLE) is negative, two replicas driven by the same noisy force show exactly the same dynamics and thus respond to this force reliably. In contrast, in the case of a positive largest LE, the states of the two replicas differ. In the context of noisy forcing, one often speaks about the synchronization of replicas by common noise~\cite{Pikovsky-84a,Pikovsky-92c,Pikovsky-Rosenblum-Kurths-01,Teramae-Tanaka-04}. However,
the same concept applies to a chaotic driving force. A reliable response of a nonlinear system to chaotic driving corresponds to a generalized synchronization phenomenon~\cite{Rulkov-Sushchik-Tsimring-Abarbanel-95}; it is also best characterized by following two replicas of the driven system. Replicas are also used in the Pecora-Carroll synchronization setup~\cite{Pecora-Carroll-90}.

In this Letter, we apply the concepts of reliability and anti-reliability to the internal dynamics of deterministic (particularly oscillatory) networks.
Complex networks often demonstrate chaotic behavior. The question we address is the following: in such a regime, does a particular unit (or, more generally, a subgroup of the units) reliably follow the driving from the other units? As discussed above, to answer this question, we have to make a replica of the unit and see if this replica has the same dynamics as the prototype unit. In the following, we consider a range of typical models and demonstrate that in some of them, all units are reliable, while in other models, some units are reliable and some are not. We establish a relation between reliability and correlations for the popular Kuramoto model of globally coupled phase oscillators, which can be interpreted as a dissipation-fluctuation relation. Furthermore, we extend the notion of internal reliability to replicated sub-networks. 

\textit{Reliability of one unit.}
We first formulate the conditions for internal reliability in a pretty general form and then consider particular examples.
Consider a set $k=1,2,\ldots,N$ of units described by variables $\mathbf{x}_k(t)$ which have their own dynamics $\sim\mathbf{F}_k$ and are coupled:
\begin{equation}
\dot{\mathbf{x}}_k=\mathbf{F}_k(\mathbf{x}_k)+\mathbf{H}_k(\mathbf{x}_k;\mathbf{x}_1,\ldots,\mathbf{x}_{k-1},\mathbf{x}_{k+1},\ldots,\mathbf{x}_N)
\label{eq1}
\end{equation}
We prepare a replica of the unit $\mathbf{x}_k$ (below, we call the replicated unit  ``prototype'') that receives the same input from other elements as the unit $\mathbf{x}_k$:
\begin{equation}
\dot{\mathbf{y}}_k=\mathbf{F}_k(\mathbf{y}_k)+\mathbf{H}_k(\mathbf{y}_k;\mathbf{x}_1,\ldots,\mathbf{x}_{k-1},\mathbf{x}_{k+1},\ldots,\mathbf{x}_N)
\label{eq2}
\end{equation}
Of course, if $\mathbf{y}_k(0)=\mathbf{x}_k(0)$, then the states of the prototype and its replica coincide for all future times. For a small deviation $\mathbf{v}_k=\mathbf{y}_k-\mathbf{x}_k$, we obtain, by subtracting \eqref{eq1} from \eqref{eq2} and linearizing, the linear system
\begin{equation}
\dot{\mathbf{v}}_k=\left(\frac{\partial \mathbf{F}_k}{\partial \mathbf{x}_k}+\frac{\partial \mathbf{H}_k}{\partial \mathbf{x}_k}\right)\mathbf{v}_k\;.
\label{eq3}
\end{equation}
Solution of this linear equation asymptotically grows, $|\mathbf{v}_k|\sim\exp[\lambda_k t]$, where $\lambda_k$ is the maximal transversal Lyapunov exponent (TLE) of \eqref{eq3}. Positive $\lambda_k$ means that the replica diverges from the prototype, and thus the unit $k$ is anti-reliable; negative $\lambda_k$ means reliability: the replica converges and eventually $\mathbf{y}_k(t)=\mathbf{x}_k(t)$ at large times.

As the first relatively trivial example, let us consider a very popular model of a recurrent neural network~\cite{Sompolinsky88}. Here, $x_k$ is a scalar, and the dynamics reads
\begin{equation}
\dot{x}_k=-x_k+\sum_{j\neq k} J_{kj}\tanh (x_j).
\label{eq4}
\end{equation}
Because the interactions do not depend on the driven unit, we have $\dot{v}_k=-v_k$, which means that all TLEs are $-1$ and complete reliability of all units. Below, we will see that replicated sub-networks of large enough size become anti-reliable.

\textit{Kuramoto model.}
Below, we will demonstrate that the properties of internal reliability are nontrivial in oscillatory networks.   
We start with a classical Kuramoto model of $N$ coupled oscillators with natural frequencies, $\omega_k$, distributed according to density $g(\omega)$:
\begin{equation}
\dot{x}_k=\omega_k+\frac{\varepsilon}{N}\sum_{j\neq k} \sin (x_j - x_k),\quad k=1,\ldots,N,
\label{eq5}
\end{equation}
where $\varepsilon$ is the coupling strength. 
In this case, equation \eqref{eq3} reads: $\dot{v}_k = - \frac{\varepsilon}{N} \sum_{j\neq k} \cos(x_j -x_k)v_k$ and the TLE for unit $k$ is:
\begin{equation}
\lambda_k = - \frac{\varepsilon}{N} \langle \sum_{j\neq k} \cos(x_j -x_k)\rangle.
\label{eq6}
\end{equation}
In the thermodynamic limit $N\to\infty$, the mean fields that enter \eqref{eq5} and \eqref{eq6} are time-independent or time-periodic, and thus the TLEs either vanish or are negative. However, the dynamics of finite ensembles is nontrivial.
In particular, the global LLE of the Kuramoto system prior to the synchronization transition is positive for finite $N$, while it tends to $0$ in the thermodynamic limit $N\to\infty$~\cite{PhysRevE.71.065201,PhysRevE.97.012203}. As a consequence, the mean fields that enter \eqref{eq5} and \eqref{eq6} fluctuate, and the TLEs can, in general, be positive. 

We thus focus our analysis on small system sizes. However, when $N$ is small, generating sampled frequencies $\omega_k$ according to a specific distribution density poses a challenge, as i.i.d. sampling can exhibit significant fluctuations in the frequency spacing. To follow the macroscopic shape of $g(\omega)$ and to avoid the burden of analyzing different samplings, we select the set of frequencies $\omega_k, k=1,\dots,N$, via a regular sampling, such that $G(\omega_k) =(k-0.5)/N$ where $G(\omega)=\int_{-\infty}^\omega g(s) ds$ is the cumulative distribution (cf.~\cite{hong2015finite,PhysRevE.97.012203,peter2018transition}). 

\begin{figure}[!htb]
    \includegraphics[width=0.48\textwidth]{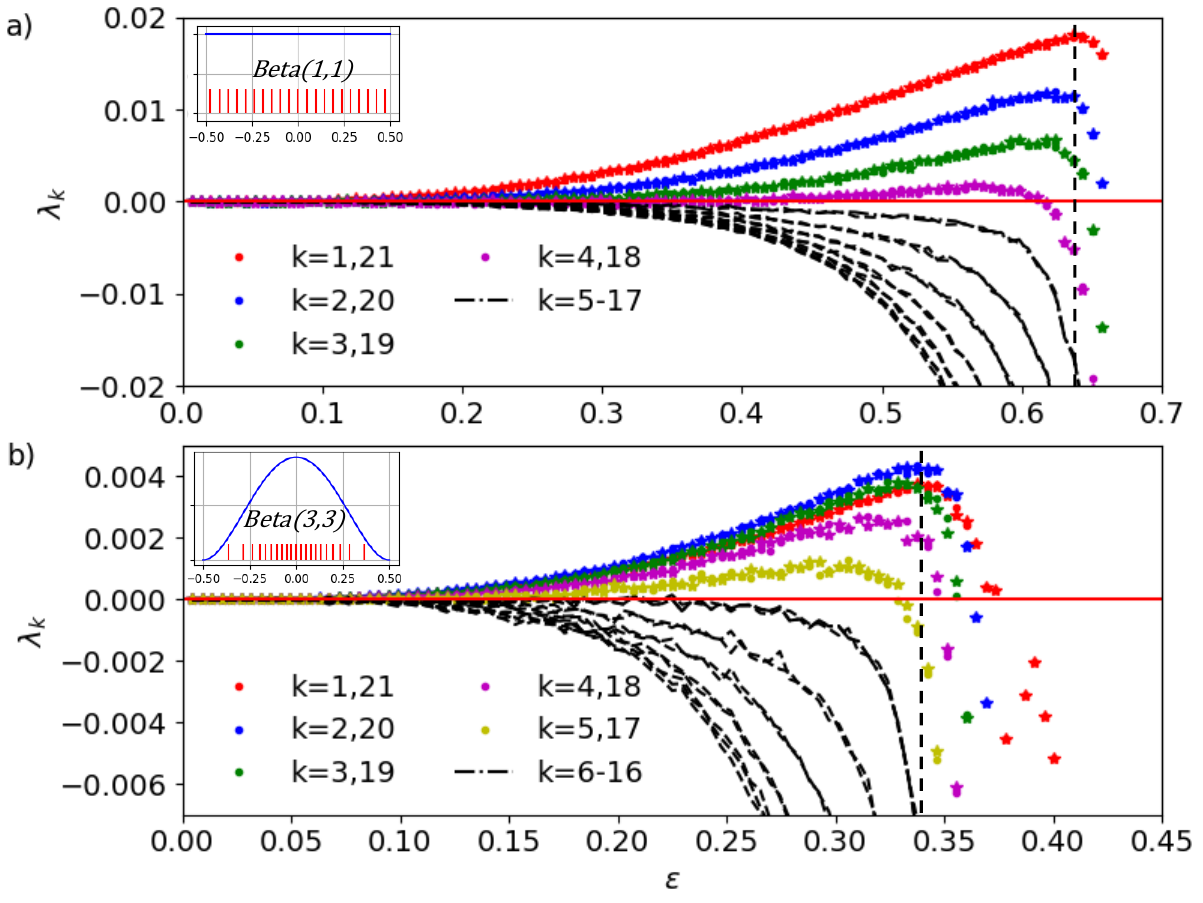}
    \caption{Transversal LEs of all Kuramoto units versus coupling strength $\varepsilon$ for $N=21$. $k$ denotes the oscillator index. 
    Units that are always reliable are shown with black dashed lines, while those anti-reliable (positive TLEs) in some range are marked by colored dots if $\omega_k>0$, and stars of the same color for symmetric units with $\omega_k<0$.
    In panel (a), frequencies are uniformly distributed (i.e., $B(\omega;1)$). In panel (b), frequencies are distributed according to $B(\omega;3)$. The insets show the shape of $g(\omega)$: the vertical red lines are the actual frequencies of the oscillators (regularly sampled, as explained in the main text). The vertical dashed lines mark the theoretical synchronization threshold (in the thermodynamic limit) $\varepsilon_c$.}
    \label{fig:img1}
\end{figure}

In Fig.~\ref{fig:img1} we show the TLEs for all oscillators in an ensemble of $N=21$ units vs. the coupling strength $\varepsilon$, for two representative distributions of the frequencies. Both are given by a symmetrized Beta distribution on the interval $[-0.5,0.5]$, $B(\omega;a) = \frac{\Gamma(2a)}{\Gamma^2(a)} [(\omega-0.5)(0.5-\omega)]^{a-1}$; $B(\omega;1)$ is a uniform distribution, and $B(\omega;3)$ is a bell-shaped one. 

One can see that all TLEs tend to zero for $\varepsilon\to 0$, but at finite coupling strengths, some units have positive TLEs and are thus anti-reliable, while TLEs of other reliable units are negative. At large $\varepsilon$, all units become reliable due to the synchronization onset (the critical coupling is well-defined $\varepsilon_c = \frac{2}{\pi g(0)}$ in the thermodynamic limit, but is slightly shifted and not so well-defined for finite $N$).

Remarkably, reliable and anti-reliable units coexist across all the intermediate range of coupling strengths, with positive $\lambda_k$ peaking near the synchronization transition. This behavior is qualitatively unaffected by the shape of the distribution $g(\omega)$, as illustrated by the two panels in Fig. \ref{fig:img1}.

\begin{figure}[!htb]
    \includegraphics[width=0.48\textwidth]{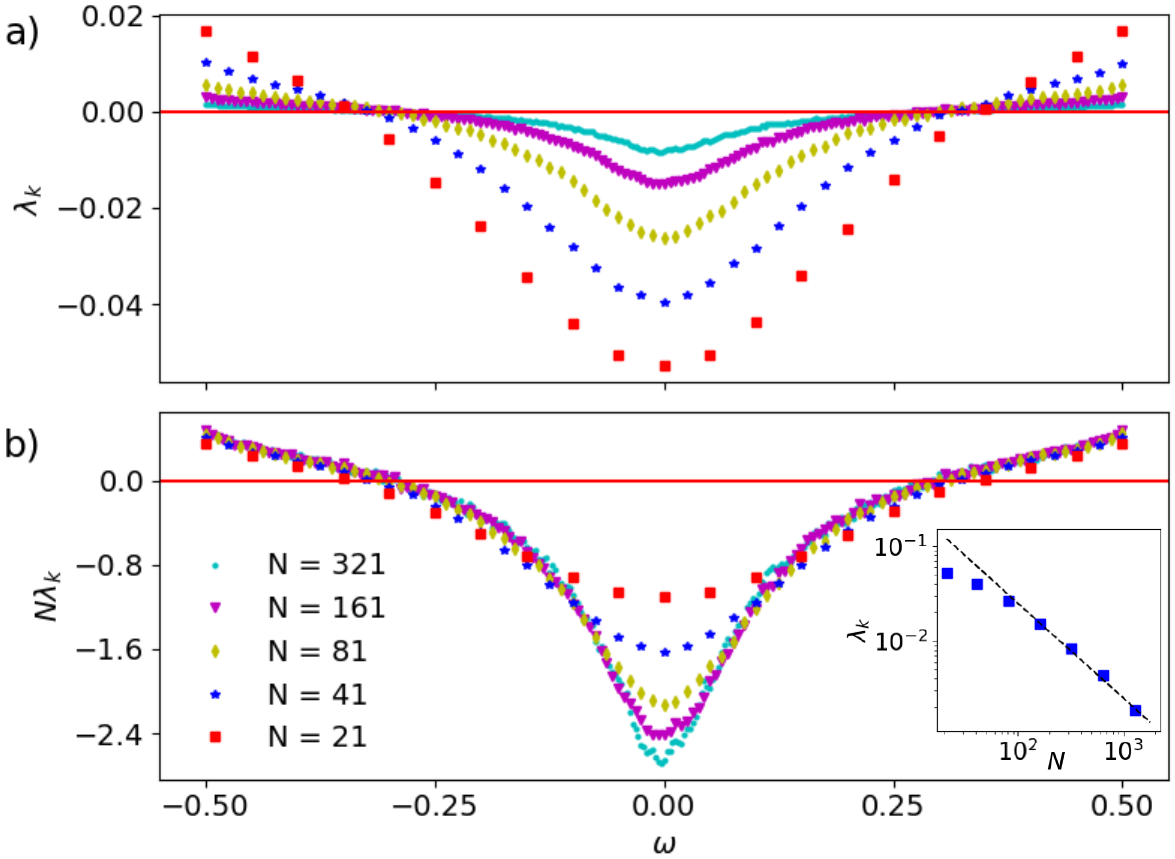}
    \caption{Panel (a): Dependence of reliability of Kuramoto units on their natural frequency ($\omega$) below the synchronization transition ($\varepsilon=0.6$, $\varepsilon_c\approx0.64$) and different ensemble size, $N=21,41,81,161,321$. Natural frequencies $\omega_k$ are uniformly distributed and equally spaced. Panel (b): The same data as in panel (a) but scaled with $N$. The inset reports the scaling of the Lyapunov exponent of the unit with $\omega=0$ (log-log scale, dashed line: $\lambda \propto 1/N)$}
    \label{fig:img2}
\end{figure}

For fixed coupling strength $\varepsilon$, reliability depends on the oscillator's natural frequency, as shown in figure \ref{fig:img2}. In particular, reliability is mostly strong (minimum of TLE) for those units whose frequency lies at the center of the distribution $g(\omega)$ and decreases toward the distribution's tails, where oscillators are anti-reliable. Comparison of different ensemble sizes shows that the magnitude of the TLE scales as $1/N$ (Fig. \ref{fig:img1} (b)). The fraction of reliable and anti-reliable oscillators in this setup does not depend on the system size, with approximately 30$\%$ being anti-reliable.

\begin{figure}[!htb]
    \centering
    \includegraphics[width=\columnwidth]{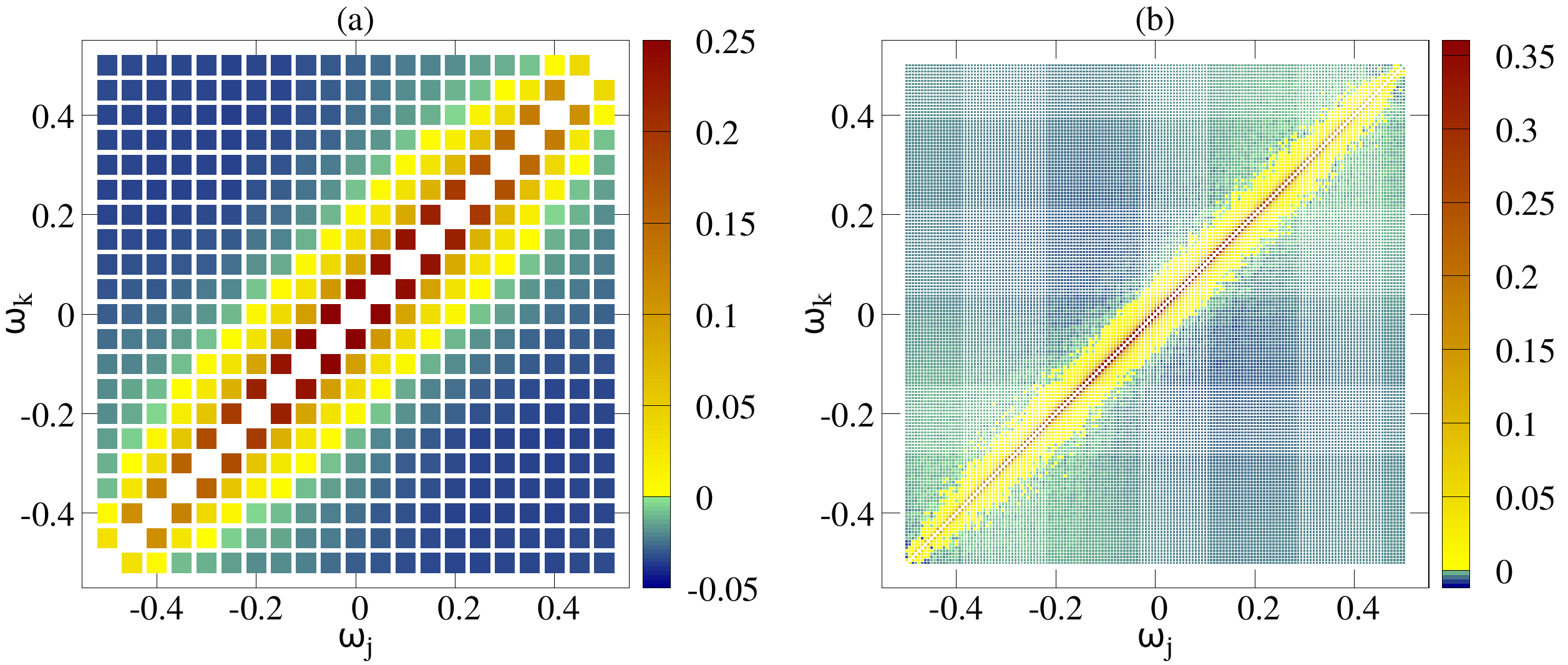}
    \caption{Correlation matrices $c_{jk}$ for the Kuramoto ensembles with a uniform distribution of natural frequencies and $N=21$ (panel (a)) and $N=161$ (panel (b)), for $\e=0.5$. To better contrast positive and negative correlations, a yellow-red palette is used for positive correlations and a green-blue one for negative values.}
    \label{fig:cor}
\end{figure}

\textit{Fluctuation-dissipation relation.}
The expression \eqref{eq6} for the TLE of the Kuramoto model allows for an interesting link to the fluctuation properties of the dynamics. One can define the pairwise correlations between the phase oscillators as $c_{jk}=c_{kj}=\langle\cos(x_j-x_k)\rangle $ (cf.~\cite{gomez2007paths,schroder2017universal,Pikovsky-Bagnoli-24}). This allows for representing the TLE \eqref{eq6} as
\begin{equation}
    \lambda_k=-\frac{\e}{N}\sum_{j\neq k} c_{jk}\;.
    \label{eq:fdr}
\end{equation}
This equality can be interpreted as a fluctuation-dissipation relation because it connects the properties of fluctuations (correlations) to the stability properties of replica-reliability (TLE). We illustrate correlations $c_{jk}$ in the Kuramoto ensemble with a uniform distribution of frequencies in Fig.~\ref{fig:cor}. 

The oscillators with close frequencies (along the diagonal) are positively correlated, although this positive correlation is small for peripheral units. Examination of different system sizes $N$ from 21 to 321 shows that the number of positively correlated neighbors is around $10-15$ and practically $N$-independent, and the values of maximal positive correlations also do not depend on $N$. In contradistinction, oscillators with significantly different frequencies are anti-correlated ($c_{jk}<0$), and the values scale as $|c_{jk}|\sim N^{-1}$. Thus, the TLE, which is according to \eqref{eq:fdr} a sum of the corresponding column/raw, is negative for central units and positive for peripheral ones, in accordance with Fig.~\ref{fig:img2}.

Remarkably, the sum of all TLEs can be expressed in another fluctuation-dissipation relation via the variance of the complex Kuramoto order parameter $Z=N^{-1}\sum_k \exp[ix_k]$. Indeed, direct calculation yields, with the account of \eqref{eq:fdr},
\begin{equation}
\begin{gathered}
\langle |Z|^2\rangle=N^{-2}\sum_{k,j}\left\langle e^{i(x_j-x_k)}\right\rangle=\\=N^{-2}\left(N+\sum_{k,j\neq k}c_{jk}\right)=N^{-1}\left(1-\e^{-1}\sum_k\lambda_k\right)\;.
\end{gathered}
\label{eq:leop}
\end{equation}
This relation is exact for any ensemble size $N$.

\textit{Other models of coupled oscillators.}We have explored different models of globally coupled oscillators that are similar to the Kuramoto model \eqref{eq5}: phase oscillators coupled via Winfree-type coupling terms~\cite{Winfree-67}; Stuart-Landau oscillators~\cite{Kuramoto-84}, coupled rotators (``oscillators with inertia'') described by second-order phase equations~\cite{Tanaka-Lichtenberg-Oishi-97}, and all of them demonstrated qualitatively the same properties as illustrated in Figs.~\ref{fig:img1},\ref{fig:img2}. Units with peripheral frequencies are anti-reliable, while those with central frequencies are reliable. However, the fluctuation-dissipation relations above are valid only for the Kuramoto model.

Furthermore, we explored several examples of random Kuramoto networks $\dot x_k=\omega_k+\e \sum_{j\neq k} A_{kj}\sin(x_j-x_k)$, with a uniform distribution of natural frequencies and randomly sampled $A_{jk}$. For both symmetric and asymmetric random matrices, the typical picture is that peripheral units are anti-reliable, and the central units are reliable. However, some central units become anti-reliable in a certain range of coupling strengths $\e$. Quantitative characterization of the internal reliability of random networks requires extensive statistical evaluations and goes beyond the scope of this Letter.

\begin{figure}[!htb]
    \centering
    \includegraphics[width=\columnwidth]{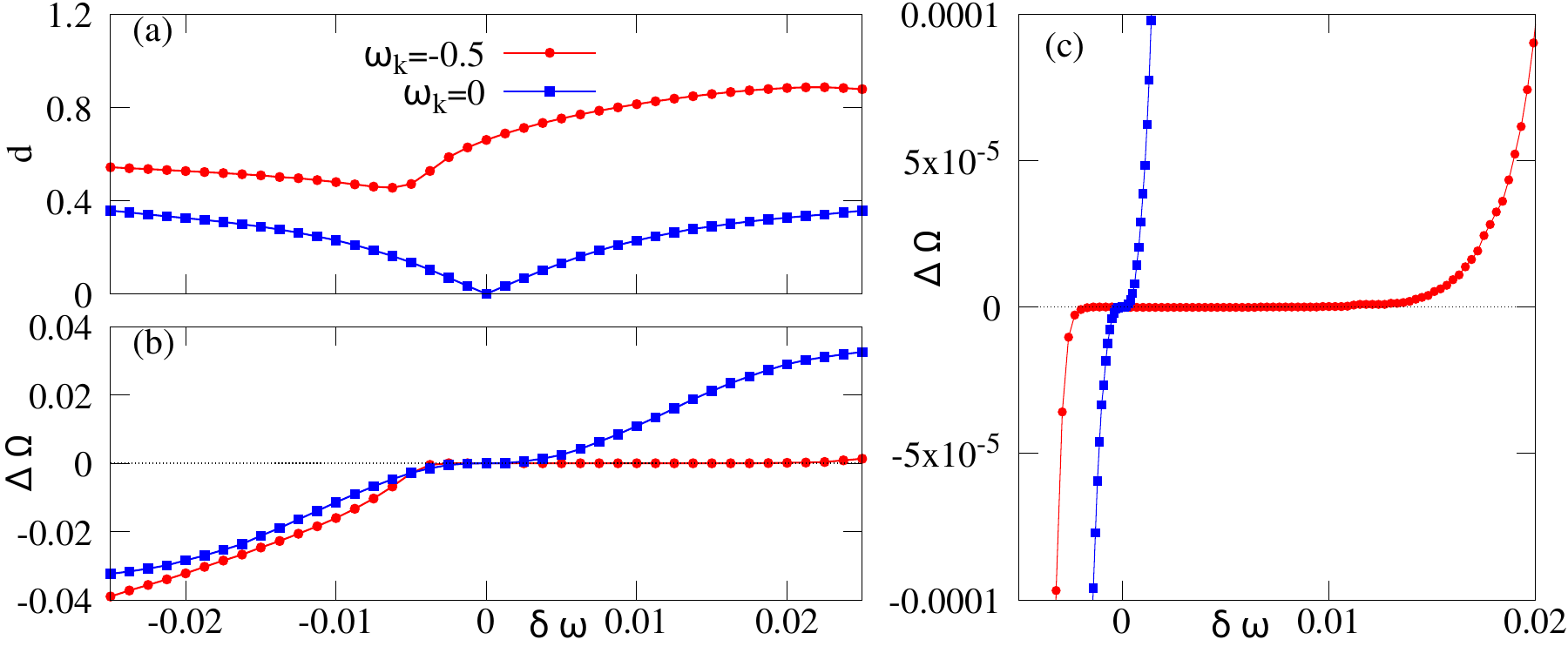}
    \caption{Difference between the phases $d$ (panel (a)) and the frequency difference $\Delta\Omega$ (panel (b)) vs $\delta\omega$ for a reliable ($\omega_k=0$, blue squares) and anti-reliable ($\omega_k=-0.5$, red circles) units. Panel (c) shows a zoomed region of panel (b) at $\Delta\Omega$. }
    \label{fig:detun1}
\end{figure}

\textit{Imperfect replicas.} 
Here, we discuss the properties of the replica dynamics if its dynamics are not identical to that of the prototype. In particular, we  take the Kuramoto ensemble \eqref{eq5}, and  replicate the unit $k$ with a detuning $\delta\omega$ in the frequency:
\begin{equation}
    \dot y_k=\omega_k+\delta\omega_k+ \frac{\varepsilon}{N}\sum_{j\neq k}\sin(x_j-y_k)\;.
    \label{eq:detun}
\end{equation}
To characterize the properties of such a replica, we calculate the time average phase difference to the prototype unit $d=\langle |\sin((x_k-y_k)/2)|\rangle$ and the difference in the observed frequencies $\Delta\Omega=\langle \dot y_k-\dot x_k\rangle$. We plot these measures as functions of the detuning $\delta\omega$ for a reliable oscillator with the central frequency $\omega_k=0$ and for an anti-reliable oscillator with $\omega_k=-0.5$ in Fig.~\ref{fig:detun1}. The behavior of the phase difference $d$ is not surprising: for the anti-reliable oscillator, it is large in the whole range of detunings, while for the reliable unit, it vanishes at $\delta\omega=0$ and grows linearly with $\delta\omega$ for small detunings. However, the behavior of frequencies is quite remarkable. For the reliable unit, the frequencies become different already for small detuning (we found that $\Delta\Omega\sim (\delta\omega)^3$). In contradistinction, for the anti-reliable unit, there is a large interval of detunings where $\Delta\Omega\approx 0$ (see panel (c) in Fig.~\ref{fig:detun1}). One can say that in this region, the replica is frequency-entrained, while the phases of the replica and the prototype remain different.

\textit{Reliability of sub-networks.} 
Here, we extend the notion of internal reliability to the sub-networks. Instead of replicating one unit like in Eq.~\eqref{eq2}, we replicate all units from a subset $\mathcal{S}$ (the non-replicated units belong to the subset $\bar{\mathcal{S}}$):
\begin{equation}
\dot{\mathbf{y}}_m=\mathbf{F}_m(\mathbf{y}_m)+\mathbf{H}_m(\mathbf{y}_m;\mathbf{x}_{l\in\bar{\mathcal{S}}},\mathbf{y}_{j\in\mathcal{S},j\neq m}),\;\; m\in\mathcal{S} 
\label{eq:subnet}
\end{equation}
The reliability is now defined via the maximal transversal Lyapunov exponents for the whole replicated subnetwork, obtained from the linear system
\begin{equation}
\dot{\mathbf{v}}_m=\left(\frac{\partial \mathbf{F}_m}{\partial \mathbf{x}_m}+\frac{\partial \mathbf{H}_m}{\partial \mathbf{x}_m}\right)\mathbf{v}_m+\sum_{j\in\mathcal{S},j\neq m} \frac{\partial \mathbf{H}_m}{\partial \mathbf{x}_j}\mathbf{v}_j,\;\;m\in\mathcal{S}\;,
\label{eq:subnetlin}
\end{equation}
which generalizes Eq.~\eqref{eq3}.

\begin{figure}[!htb]
    \includegraphics[width=\columnwidth]{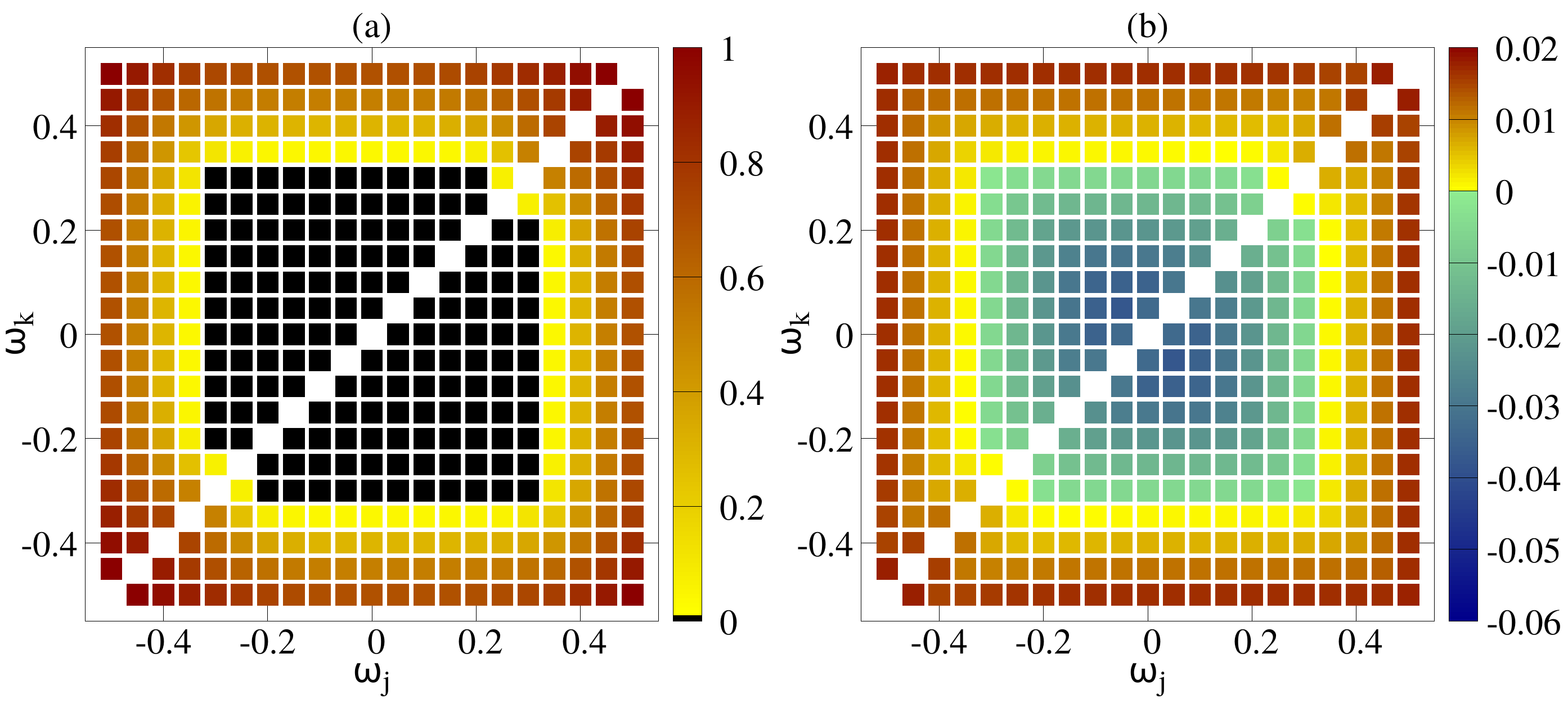}
    \caption{Kuramoto model with a uniform distribution of frequencies, $N=21$, $\e=0.6$. Panel (a): distance $d=\langle(\sin^2((x_k-y_k)/2)+\sin^2((x_j-y_j)/2))^{1/2}\rangle$ between the two prototypes and the corresponding replicas. The values $d<0.01$ are depicted with black, otherwise the color palette is used. Panel (b): TLEs for the pair of replicas calculated according to \eqref{eq:subnetlin}, here two palettes for positive and negative values are used for better contrast.}
    \label{fig:tworep}
\end{figure}

\begin{figure}[!htb]
    \includegraphics[width=1\columnwidth]{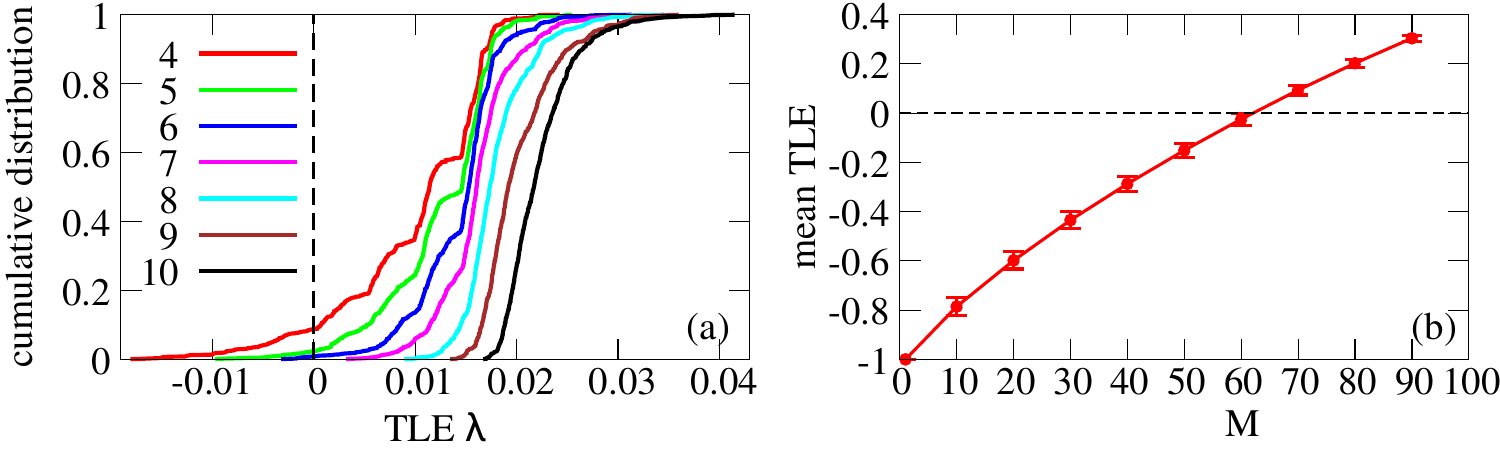}
    \caption{Panel (a): Cumulative distributions of the TLEs for the Kuramoto model of $N=21$ units \eqref{eq5}, for different sizes of the subnetworks. Panel (b): Mean values and standard deviations (depicted as error bars) of the TLEs for randomly sampled sub-networks of the recurrent neural network \eqref{eq4}.}
    \label{fig:distr}
\end{figure}

Let us first consider the Kuramoto model and different subnetworks of two replicas. Here, one can look at all combinations of two selected units for replication; the results are presented in Fig.~\ref{fig:tworep}. One can see that practically in all cases,  a pair is reliable if both units are individually reliable. This rule works only as a ``rule of thumb'' for larger replicated sub-networks. The number of possible larger sub-networks is too large to explore all of them, thus we performed a statistical analysis and found the transversal LEs for a random sample of 1000 sub-networks of sizes $4,\ldots, 10$ out of $N=21$, for the uniform distribution of natural frequencies. The cumulative distributions of the LEs are presented in Fig.~\ref{fig:distr}(a). One can see that already for sub-networks of size $4$, only  10\% of all cases are reliable, and starting from size $7$, all tested subnetworks are anti-reliable.

In the next example, we explore the reliability of sub-networks in the recurrent neural network \eqref{eq4}. As mentioned, all individual elements are reliable here with the trivial TLE equal to $-1$. To see how the LE depends on the size of the replicated sub-network, we have chosen one neural network of size $N=100$ operating in a chaotic regime and explored 1000 randomly sampled sub-networks for sizes $M=10,20,\ldots,90$. The distributions of the calculated LEs were nearly Gaussian, and we plot in Fig.~\ref{fig:distr}(b) dependencies of the mean and the standard deviation on the sub-network size. One can see that the transition is at $M\approx 62$: larger sub-networks are practically all anti-reliable, while smaller sub-networks are all reliable.



\textit{Conclusion.}
Summarizing, we explored the internal reliability properties of dynamical networks by determining the stability of particular units and sub-networks when replicated. The standard Kuramoto model of globally coupled phase oscillators shows nontrivial reliability properties, with anti-reliable peripheral and reliable central oscillators. In this model, these properties are well-defined for finite networks but disappear in the thermodynamic limit, where the acting fields are constant or periodic. 
For the recurrent neural network model, which is chaotic in the thermodynamic limit, replicas of small sub-networks are reliable. Still, starting from a certain size, the replicated sub-networks become anti-reliable.

While the concept applies to generic regular and random dynamical networks, for the globally coupled Kuramoto model, we additionally established a fluctuation-dissipation relation. This relation connects the stability exponent of internal reliability of a particular unit to its pairwise phase correlations with others. The sum of all exponents can be expressed in another fluctuation-dissipation relation via the variance of the complex Kuramoto order parameter. 

The reliability property is important for the identification of the parameters of the elements of the network. As we demonstrated, the difference of states between the replica and the prototype, if they have the same parameters, vanishes only in the case of reliability, whereas it remains finite in the case of anti-reliability.

\acknowledgments
\textit{Acknowledgments.} T.M. was funded by the European Union-NextGenerationEU-National Recovery and Resilience Plan, Mission 4 Component 2-Investment 1.5-THE-Tuscany Health Ecosystem-ECS00000017-CUP B83C22003920001.
%

\end{document}